# Two fluid cosmological models in $f(R, T)$ theory of gravity

[*]Y. S. Solanke[1], [**]Sandhya Mhaske., [+]D.D. Pawar[2], [++]V. J. Dagwal[3]

[1] Department of Mathematics, Mungasaji Maharaj Mahavidyalaya, Darwha, Yavatmal 445202, India

[2]School of Mathematical Sciences, S. R. T. M. University, Nanded, Maharashtra 431606, India

[3]Department of Mathematics, Government College of Engineering, Nagpur441108, India

Email-: [*]yadaosolanke@gmail.com ,[**]mhaske.ss924@gmail.com, [+]dypawar@yahoo.com, [++]vdagwal@gmail.com

**Abstract:**

Present work deals with the two fluid Bianchi Type-V cosmological models consisting of matter and radiating source in the $f(R, T)$ theory of gravity studied by Harko et al. (2011). In this paper, we developed a new idea about $f(R, T)$ gravity with the help of two fluids: one fluid is matter field modeling material content of the Universe and other fluid is radiation field modeling the CMB. We have determined the solution of the two fluid gravitational field equations with the systematic structure in $f(R, T)$ gravity. Here we have deliberated four types of universe such as dust universe, radiation universe, hard universe and Zel'dovich universe and also extended our work to observe the big rip and big bang singularity. We have also tested the cosmological parameters.

**Keywords:** Two fluid, $f(R,T)$ theory, Bianchi Type V space time.

1. **Introduction:**

As compare to two sequences of a single-fluid model, the cosmic evolution based on two-fluid big-bang universe appears much better. Two-fluid universe explains the change among radiation-dominated phases to a matter-dominated phase. Cosmologists have described that our present universe is expanding as well as accelerating. Ia supernovae and cosmic microwave background strongly recommend that current model is conquered by the dark energy which arises because of cosmic-accelerated expansion of the universe. Numbers of researchers have studied several aspects of two-fluid cosmological models with matter content of the universe and radiating source in Einstein's general relativity (GR) and also in alternative theory of gravity.

Cosmological models with matter and radiating source in general relativity are examined by Oli (2008). Einstein's field equation and Higher dimension cosmological universe with mixture of fluid in GR are proposed by McIntosh (1972) and Samanta (2013). Physical property of Bianchi type-II universe with matter and radiating source in general relativity are discussed by Pant and Oli (2002). Singh *et al.*(2013) have obtained the deterministic solution of two-fluid Bianchi type-V cosmological models in GR with the support of negative constant deceleration parameter. Pawar and Dagwal (2014) have explored the physical properties of mixture of fluids with tilted cosmological universe in GR and presented various types of universe such as dust universe, radiation universe, hard universe and Zel'dovich universe. Dagwal and Pawar (2018) have extended their work of tilted cosmological universe with two-fluid sources in general relativity with assistance of a variables $G$ and $\lambda$. Samanta (2013) has discussed kinematical properties and behavior of two-fluid sources with the help of variables $G$ and $\lambda$ in Einstein's field equation. Verma *et al.* (2015) are obtained planar cosmic space time with matter and radiating source in general relativity. Coley (1988) and Dunn (1989) have formulated Einstein's field equation with two-fluid sources of Bianchi type-VI0 universe. Pawar *et al.* (2015) have explored several types of universe with mesonic two-fluid sources in Lyra Geometry. Harko *et al.*(2011) have discussed of kinetic model for two-fluid in gravitational field equations. Dagwal and pawar (2020) have formulated the new idea about matter and radiating source inin $f(T)$ theory of gravity and discussed geometric property with help of MATLAB. Thermodynamics property of two-fluid sources is presented by Coley (1988) and Dunn (1989) has observed two-fluid gödel-type space time cosmological universe. Dagwal (2020) has studied mesonic two-fluid sources cosmic model in $f(T)$ theory of gravity. Solanke *et al.*(2021) have obtained accelerating cosmic model with mixture of fluids, first fluid show the perfect fluid and other show the dark energy.

Current astrophysical remark presented the expansion of the universe in an accelerated era. The observational data of supernovae type Ia1 and cosmic microwave background (CMB) have explored that our universe is expanding at an increasing rate. An amended theory of gravity can be employed to address some problems of current interest and may lead to some main modifications. Many researchers have studied various aspects of modified theory of gravity like as $f(T)$, $f(G)$, $f(R)$, $f(R,T)$ and $f(G,T)$ etc. In this work, we have studied $f(R,T)$ modified theory of gravity, where $R$ is the Ricci scalar and $T$ is the trace of the energy-momentum tensor.

Harko et al. (2011); Bertolami et al. (2008); Myrzakulov (2012); Sharif and Zubair (2012); Houndjo (2012); Adhav (2012); katore (2012); Alvarenga et al. (2013); Samanta (2013a), (2013b); Harko and Lobo (2014); Baffou(2015); Pawar and solanke(2015); Yousaf et al.(2017) Moraes and Sahoo(2017); Pradhan et al.(2021); Pacif et al. (2015), (2020); Dagwal et al. (2020) ; Kata's(2019a),(2019b) have developed different aspects of $f(R, T)$ gravity. Yousuf (2020a), (2020b) has discussed self-gravitating system in $f(R, T)$ gravity and electromagnetic field in modified theory of gravity. Dagwal and Pawar (2020a) (2020b) have discussed tilted and non-tilted universe in modified theory of gravity. Nagpalet al.

(2018) have discussed geometrical and dynamics properties of cosmological model in f(R, T) modified theory of gravity.

Motivated by above work here we have studied two-fluid Bianchi Type V cosmological models with matter and radiating source in the f(R, T) theory of gravity. We have tested the cosmological parameters.

This paper is organized as - Section 2: Metric and field equation, Section 3: Physical and geometrical Properties, Section 4: Cosmological parameters, Sub section 4.1: Look-back time-redshift, Sub section 4.2: Proper Distance, Sub section 4.3: Luminosity distance, Sub section 4.4: Angular-diameter distance, Sub section 8.5: Distance Modulus, Section 5: Results and discussion, Section 6: Conclusion.

.

**2. Metric and field equation:**

We have considered Bianchi type V space time of the form

$$ds^2 = -dt^2 + A^2 dx^2 + e^{2x}\left(B^2 dy^2 + c^2 dz^2\right), \tag{1}$$

where A, B, C are functions of cosmic time t alone.

The Einstein's field equation in $f(R,T)$ theory of gravity for the function

$$f(R,T) = R + 2f(T) \tag{2}$$

given as

$$R_{ij} - R g_{ij} = T_{ij} + 2f'T_{ij} + \left[2pf'(T) + f(T)\right] g_{ij}, \tag{3}$$

where $T_{ij}$ is the energy momentum tensor for the perfect fluid and it is given by

$$T_{ij} = T_{ij}^{(m)} + T_{ij}^{(r)}, \tag{4}$$

where $T_{ij}^{(m)}$ is energy momentum tensor for matter field having density $\rho_m$, pressure $p_m$ and $u_1^{(m)} = (0,0,0,1)$ as four velocities with $g^{ij} u_i^{(m)} u_j^{(m)} = 1$. $T_{ij}^{(r)}$ is energy momentum tensor for radiation field having density $\rho_r$, pressure $p_r$ and $u_1^{(r)} = (0,0,0,1)$ as four velocity with $g^{ij} u_i^{(r)} u_j^{(r)} = 1$ and $p_r = \frac{1}{3}\rho_r$.

Thus, with above information $T_{ij}^{(m)}$ and $T_{ij}^{(r)}$ are given by

$$T_{ij}^{(m)} = (\rho_m + p_m)u_i^{(m)}u_j^{(m)} - p_m g_{ij}, \tag{5}$$

$$T_{ij}^{(r)} = \frac{4}{3}\rho_r u_i^{(m)}u_j^{(m)} - \frac{1}{3}\rho_r g_{ij}. \tag{6}$$

We have consider

$$f(T) = \lambda T. \tag{7}$$

Equation of state

$$p_m = (\gamma - 1)\rho_m \qquad 1 \leq \gamma \leq 2 \tag{8}$$

Now, with the help of (5) and (6) the field equations (2) for metric (1) yield

$$\frac{B_{44}}{B} + \frac{C_{44}}{C} + \frac{B_4 C_4}{BC} - \frac{1}{A^2} = -p_m(1+5\lambda) - \frac{\rho_r}{3}(1+8\lambda) - \lambda\rho_m, \tag{9}$$

$$\frac{A_{44}}{A} + \frac{C_{44}}{C} + \frac{A_4 C_4}{AC} - \frac{1}{A^2} = -p_m(1+5\lambda) - \frac{\rho_r}{3}(1+8\lambda) - \lambda\rho_m, \tag{10}$$

$$\frac{A_{44}}{A} + \frac{B_{44}}{B} + \frac{A_4 B_4}{AB} - \frac{1}{A^2} = -p_m(1+5\lambda) - \frac{\rho_r}{3}(1+8\lambda) - \lambda\rho_m, \tag{11}$$

$$\frac{A_4 B_4}{AB} + \frac{B_4 C_4}{BC} + \frac{A_4 C_4}{AC} - \frac{3}{A^2} = p_m(2+7\lambda) + \rho_m(1+3\lambda) - \frac{\rho_r}{3}(5+17\lambda), \tag{12}$$

$$2\frac{A_4}{A} - \frac{B_4}{B} - \frac{C_4}{C} = 0. \tag{13}$$

Here the suffix 4 to the field variable represents the differentiation with respect to cosmic time $t$.

From equation (13) we have

$$A^2 = lBC,$$

Here $l$ is constant of integration.

Without any loss of generality we can take this constant as unity, so we have,

$$A^2 = BC \tag{14}$$

Equations (9) to (13) are five independent equations with 6 unknowns. In order to find the determinate solution we use the fact that scalar expansion $\theta$ is proportional to shear scalar $\sigma^2$, hence we get

$$B = C^k. \tag{15}$$

Where $k \neq 1$ is constant which preserves the isotropic character of the space time.

Subtracting (10) from (11)

$$\frac{B_{44}}{B} - \frac{C_{44}}{C} + \frac{A_4 B_4}{AB} - \frac{A_4 C_4}{AC} = 0. \tag{16}$$

Using (14) and (15) for solving (16), we get

$$A = b_1 T^{\frac{1}{3}}, \ B = b_2 T^{\frac{2k}{3k+3}}, \ C = b_3 T^{\frac{2}{3k+3}}, \tag{17}$$

where $b_1, b_2, b_3$ are constants and $T = t + d$ where $d$ is constant of integration.

Thus, line element (2.1) reduced to

$$ds^2 = -dT^2 + b_1^2 T^{\frac{2}{3}} dx^2 + e^{2x} \left( b_2^2 T^{\frac{4k}{3k+3}} dy^2 + b_3^2 T^{\frac{4}{3k+3}} dz^2 \right)$$

$$ds^2 = -dT^2 + b_1^2 T^{\frac{2}{3}} dx^2 + e^{2x} T^{\frac{4}{3k+3}} \left( b_2^2 T^k dy^2 + b_3^2 dz^2 \right) \tag{18}$$

### 3. Physical and geometrical properties

The directional Hubble parameter in the direction of *x, y* and *z* are respectively given by

$$H_1 = \frac{1}{3T}, \ H_2 = \frac{2k}{3k+3}\frac{1}{T}, \ H_3 = \frac{2}{3k+3}\frac{1}{T}. \tag{19}$$

The Hubble parameter *H* is

$$H = \frac{1}{3T}. \tag{20}$$

Average scale factor is given by

$$H = \frac{\dot{R}}{R} = \frac{1}{3T}.$$

Integrating

$$R = c'^2 T^{\frac{1}{3}}, \qquad (21)$$

where $c'$ is constant of integration.

The energy conservation equation $T_{j,i}^{i} = 0$ leads to two equations for matter density and radiation density

$$(7\lambda\gamma - 4\lambda + 2\gamma - 1)\rho_{m_4} + (H_1 + H_2 + H_3)(12\lambda\gamma - 10\lambda + 3\gamma - 2)\rho_m = 0, \qquad (22)$$

$$(5 + 17\lambda)\rho_{r_4} + (H_1 + H_2 + H_3)(4 + 9\lambda)\rho_r = 0. \qquad (23)$$

From (18), (19) and (21) we get

$$\rho_m = \eta_1 T^{-\left(\frac{12\lambda\gamma - 10\lambda + 3\gamma - 2}{7\lambda\gamma - 4\lambda + 2\gamma + 1}\right)}, \qquad (24)$$

$$\rho_r = \eta_2 T^{-\left(\frac{4 + 9\lambda}{5 + 17\lambda}\right)}, \qquad (25)$$

where $\eta_1$ and $\eta_2$ are constants of integration.

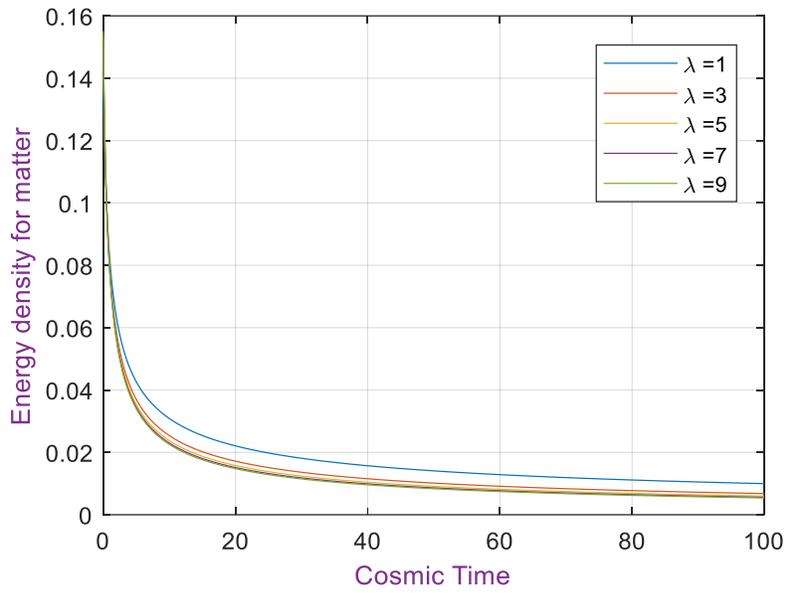

Fig.1a Energy density for matter against cosmic time *t* (Gyr)

with $\eta_1 = 0.1$, $d = 0.5$, $\gamma = 1$ and for the values of $\lambda$ =1, 3, 5, 7, 9.

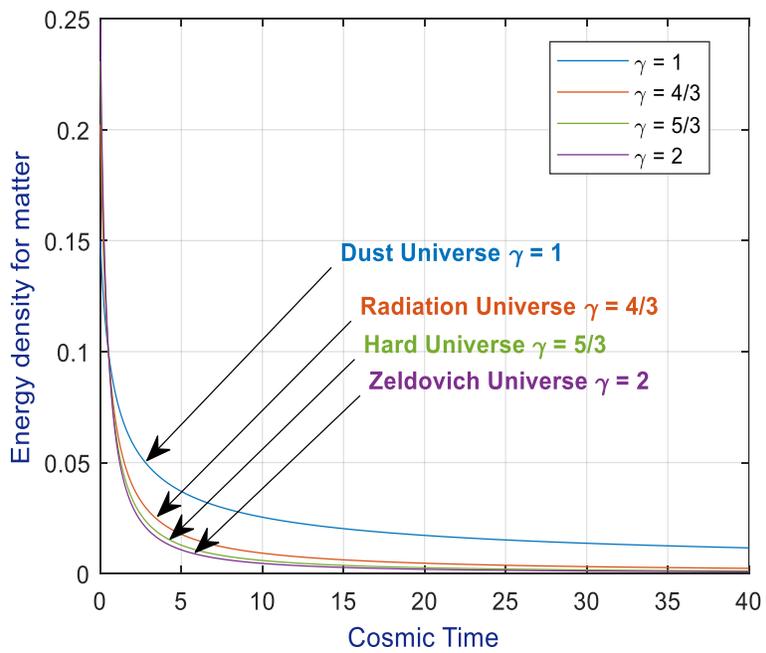

Fig.1b Energy density for matter against cosmic time *t* (Gyr)

With $\eta_1 = 0.1$, $d = 0.5$, $\lambda = 3$ and for the values of $\gamma$ =1, 4/3, 5/3, 2.

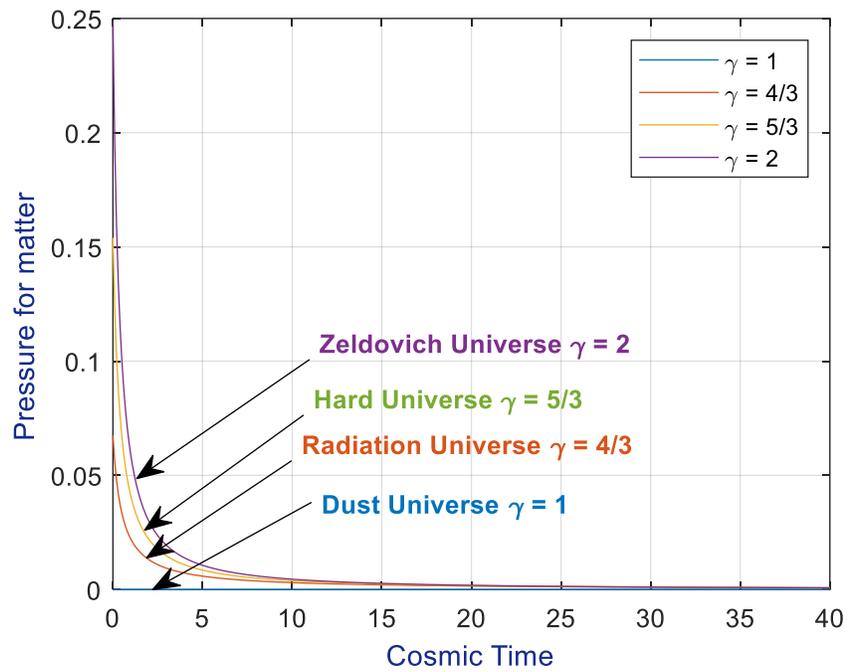

Fig.1c Pressure for matter against cosmic time $t$ (Gyr)

with $\eta_1 = 0.1$, $d = 0.5$, $\lambda = 3$ and for the values of $\gamma = 1, 4/3, 5/3, 2$.

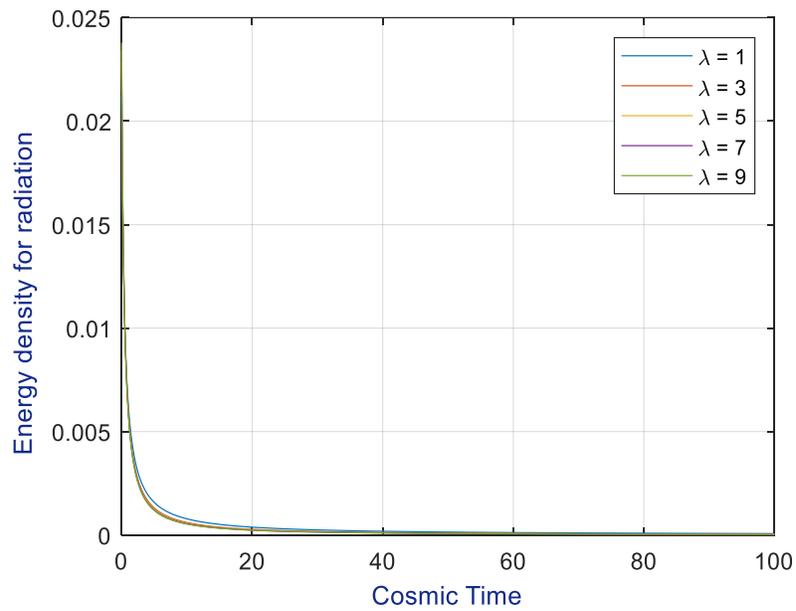

Fig.2 Energy density for radiation against cosmic time $t$ (Gyr)

with $\eta_2 = 0.1$, $d = 0.5$, $\gamma = 1$ and for the values of $\lambda$ =1, 3, 5, 7, 9.

The density parameters are

$$\Omega_m = 3\eta_1 T^{-\left(\frac{12\lambda\gamma-10\lambda+3\gamma-2}{7\lambda\gamma-4\lambda+2\gamma+1}\right)+2}, \tag{26}$$

$$\Omega_r = 3\eta_2 T^{-\left(\frac{4+9\lambda}{5+17\lambda}\right)+2} \tag{27}$$

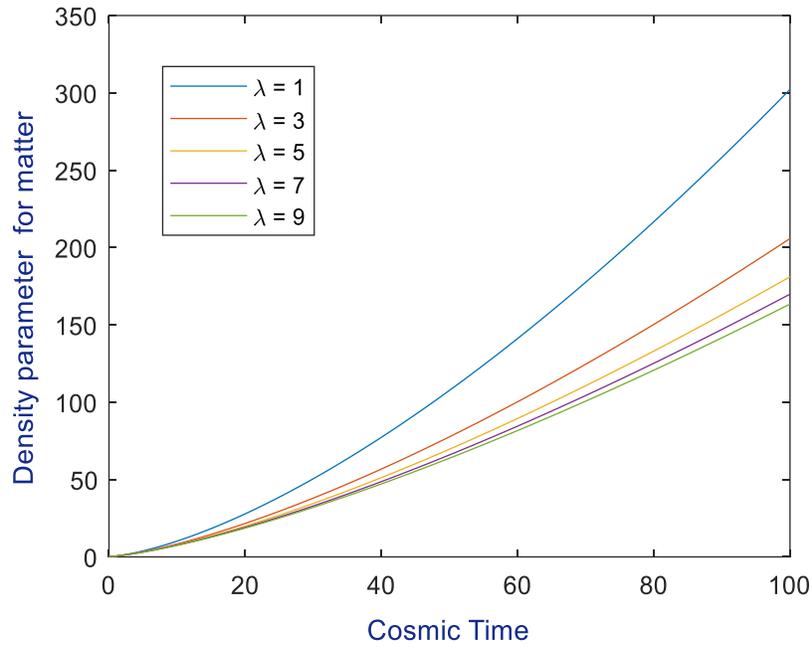

Fig.3a Density parameter for matter against cosmic time $t$ (Gyr)

with $\eta_1 = 0.1$, $d = 0.5$, $\gamma = 1$ and for the values of $\lambda$ =1, 3, 5, 7, 9.

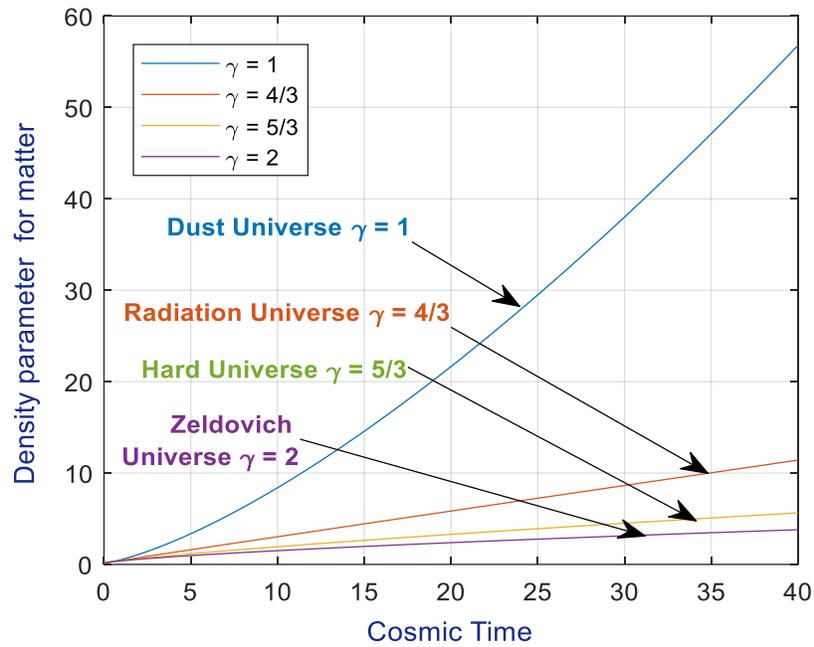

Fig.3b Density parameter for matter against cosmic time $t$ (Gyr)

with $\eta_1 = 0.1$, $d = 0.5$, $\lambda = 3$ and different values of $\gamma$ =1, 4/3, 5/3, 2.

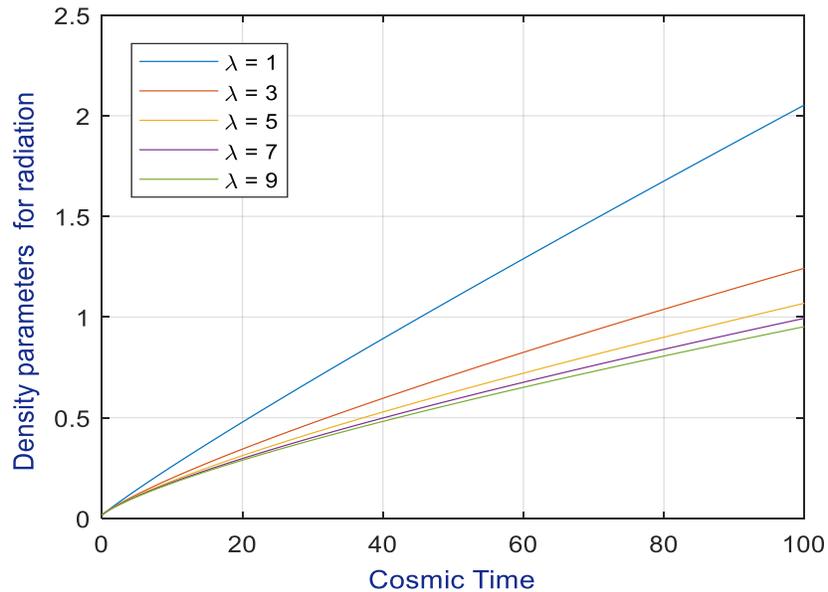

Fig.4 Density parameter for radiation against cosmic time $t$ (Gyr)

with $\eta_2 = 0.1$, $d = 0.5$, $\gamma = 1$ and for the values of $\lambda$ =1, 3, 5, 7, 9.

The total density parameter for the derived model is

$$\Omega = 3\left[\eta_1 T^{-\left(\frac{12\lambda\gamma-10\lambda+3\gamma-2}{7\lambda\gamma-4\lambda+2\gamma+1}\right)+2} + \eta_2 T^{-\left(\frac{4+9\lambda}{5+17\lambda}\right)+2}\right] \qquad (28)$$

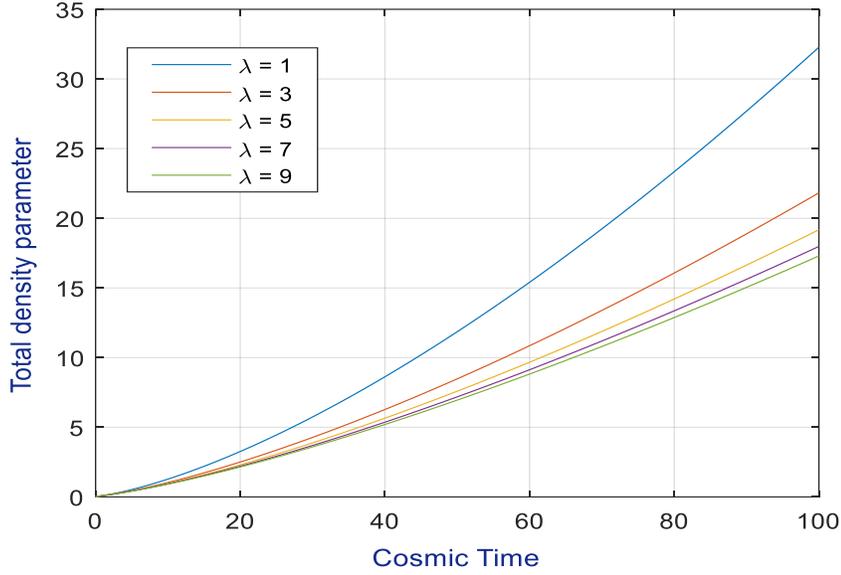

Fig.5a Total density parameter against the cosmic time $t$ (Gyr)

with $\eta_1 = 0.1$, $\eta_2 = 0.1$, $d = 0.5$, $\gamma = 1$ and for the values of $\lambda$ =1, 3, 5, 7, 9.

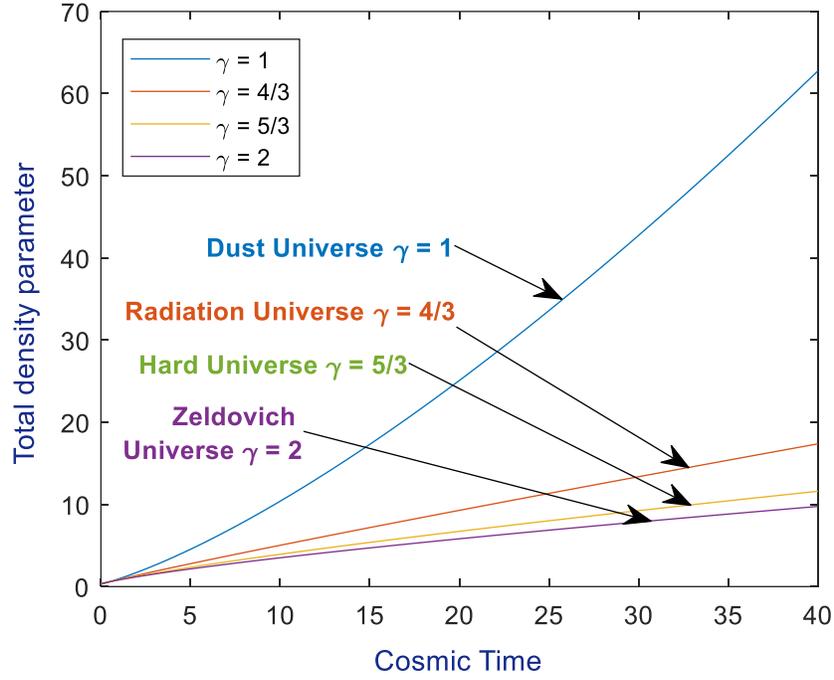

Fig.5b Total density parameter for matterverses cosmic time $t$ (Gyr)

with $\eta_1 = 0.1$, $\eta_2 = 0.1$, $d = 0.5$, $\lambda = 3$ and for the values of $\gamma$ =1, 4/3, 5/3, 2.

Anisotropic parameter is given by

$$A_m = \frac{2}{3}\left(\frac{k-1}{k+1}\right)^2. \qquad (29)$$

The scalar expansion is

$$\theta = \frac{1}{T} \qquad (30)$$

andshear scalar is given by

$$\sigma^2 = \frac{(k-1)^2}{9(k+1)^2 T^2} \qquad (31)$$

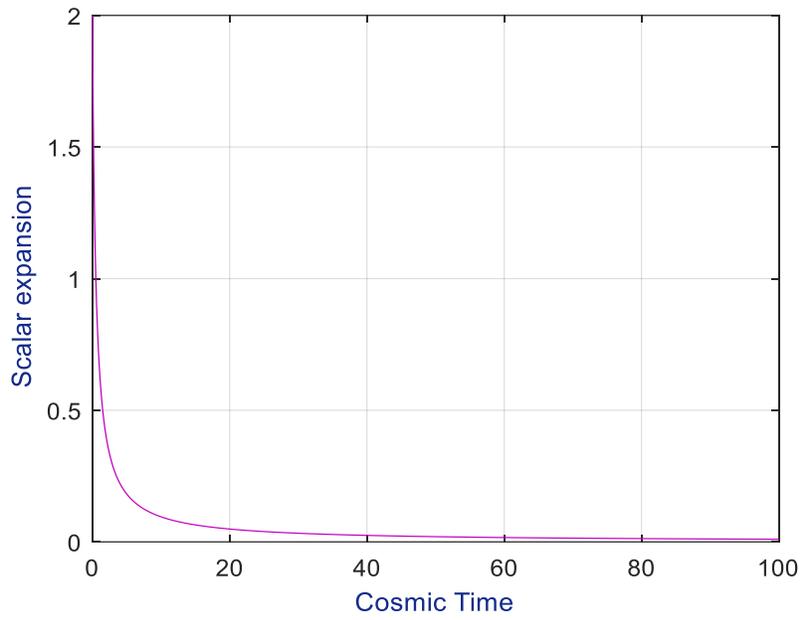

Fig.6 Conduct scalar expansion against time $t$ (Gyr) with $d = 0.5$

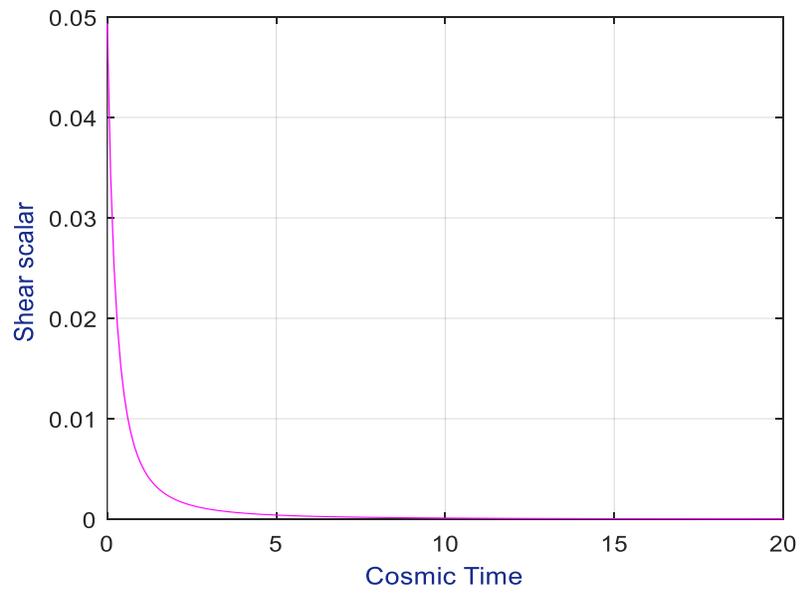

Fig.7 Conduct of shear scalar against time $t$ (Gyr)

for $d = 0.5$ and $k = 2.5$

The deceleration parameter is found to be

$$q = 2 \tag{32}$$

The spatial volume is

$$V = b'T, \tag{33}$$

where b' is a integration constant.

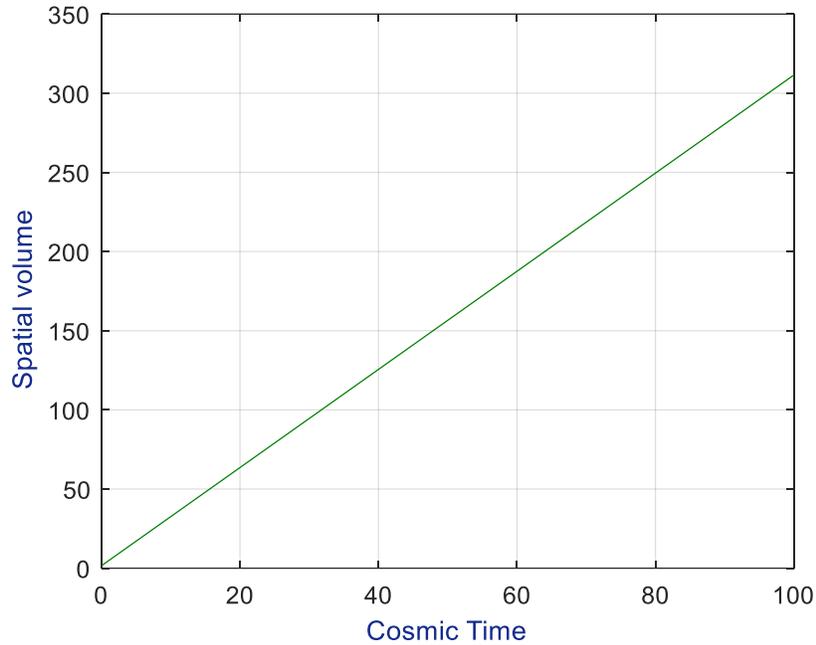

Fig.8 Conduct of spatial volume with respect to cosmic time $t$ (Gyr)

with $d = 0.5$ and $b' = 3.1$

The profile of spatial volume verses cosmic time is represented by figure 8 by taking the values $d = 0.5$ and $b' = 3.1$. The spatial volume is an increasing function of time and it approaches to infinity with the evolution of time. Equation (33) represents that the spatial volume is diverges for $T \to \infty$. It stops at $T \to 0$. From figure 8, the spatial volume is increasing with respect to cosmic time.

## 4. Cosmological parameters
The cosmological parameters are given by
### 4.1 Look back time-red-shift:

Cosmological red-shift $z$ is directly proportional to size of Universe i. e. scale factor $R(t)$

$$1 + z = \frac{R(t_0)}{R(t)} = \frac{R_0}{R}. \tag{34}$$

Suffix '0' refers to present epoch. Here $R_0$ is the size of universe at the time the light from object is observed and $R$ is the size of Universe at the time it was emitted. i.e. scale factor.

Look back $t_L$ time is given by

$$t_L = t_0 - t(z) = \int_R^{R_0} \frac{dR}{\dot{R}} \tag{35}$$

From equations (21) and (34)

$$H_0(t_0 - t) = \frac{1}{(q+1)}\left[1 - (1+z)^{-3}\right], \tag{36}$$

$$H_0(t_0 - t) = z - 2z^2 + ....., \tag{37}$$

where $H_0$ is the Hubble constant at present and $(q+1)=3$. The value of Hubble constant $H_0$ is lies between $50-100 \text{ km s}^{-1}\text{ Mpc}^{-1}$.

Taking $z \to \infty$ in (36) we get

$$H_0(t_0 - t) = \frac{1}{(q+1)},$$

$$t_L = (t_0 - t) = \frac{1}{(q+1)H_0} = \frac{H_0^{-1}}{(q+1)},$$

$$t_L = \frac{H_0^{-1}}{(q+1)}. \tag{38}$$

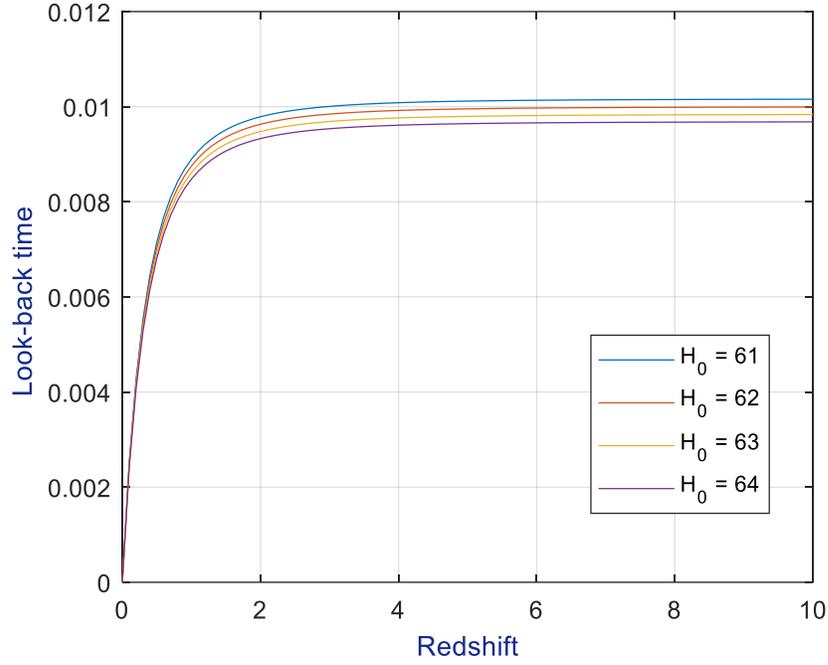

Fig.9 Conduct of look-back time $t_L$ with respect to redshift $z$

with $q = -0.38$ and for the value of $H_0$ =61, 62, 63, 64.

The graph of $t_L$ against $z$ is presented by Figure 9 by taking $q = -0.38$ and different values of $H_0$.

**4.2 Proper distance:**

The proper distance $d(z)$ given by

$$d(z) = r_1 R_0, \tag{39}$$

where $r_1 = \int_t^{t_0} \frac{dt}{R}$, (40)

$$d(z) = \frac{9}{qc'^2} H_0^{-1} \left[1 - (1+z)^{-2}\right]. \tag{41}$$

When $z \to \infty$, the proper distance $d(z)$ is

$$d(z) = \frac{9}{qc'^2} H_0^{-1} \tag{42}$$

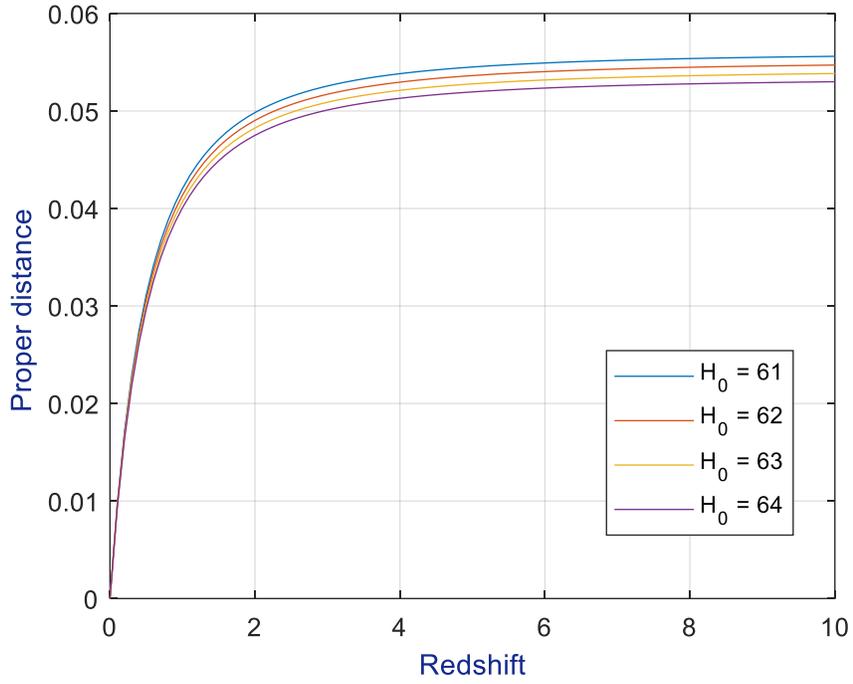

Fig.10 Conduct of proper distance *d(z)* with respect to *z*

with $q = -0.38$, $c' = -1$ and for the values of $H_0$ =61, 62, 63, 64.

The graph of proper distance *d(z)* *verses* redshift *z* is presented by figure 10 by taking the values $q = -0.38$, $c' = -1$ and different values of $H_0$.

**4.3 Luminosity distance:**

Luminosity distance is given by

$$d_L = r_1(z)R_0(1+z) = d(z)(1+z) \tag{43}$$

From equation (41)

$$d_L = \frac{9}{qc'^2} H_0^{-1}\left[1-(1+z)^{-2}\right](1+z) \tag{44}$$

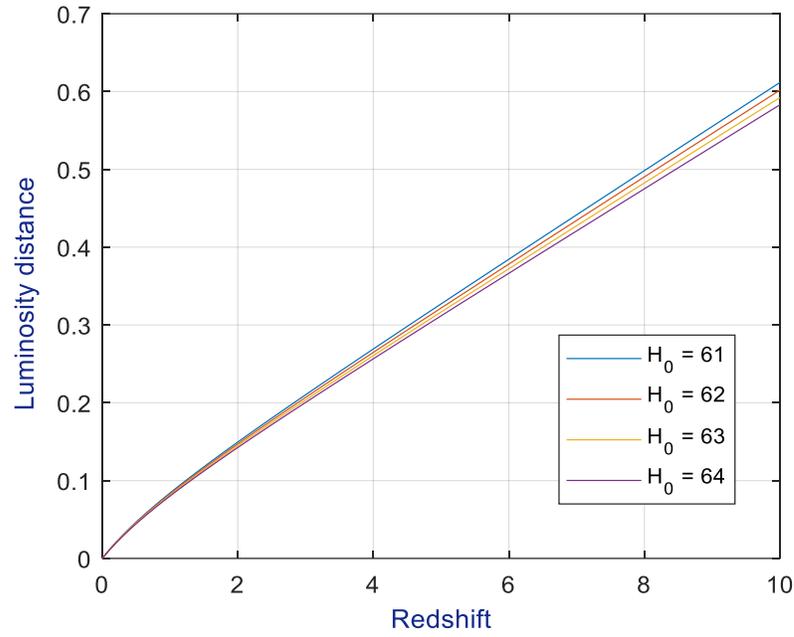

Fig.11 Conduct of luminosity distance with respect to redshift $z$

with $q = -0.38$, $c' = -1$ and for the values of $H_0$ =61, 62, 63, 64.

The graph of luminosity distance with respective $z$ is presented by figure 11 by taking the ranges $q = -0.38$, $c' = -1$ and different values of $H_0$.

**4.4 Angular diameter distance**

The angular diameter distance is given by ratio of an object's physical transverse size to its angular size. Angular diameter distance is given by

$$d_A = d(z)(1+z)^{-1} = d_L(1+z)^{-2} \tag{45}$$

$$d_A = \frac{9}{qc'^2} H_0^{-1}\left[1-(1+z)^{-2}\right](1+z)^{-1} \tag{46}$$

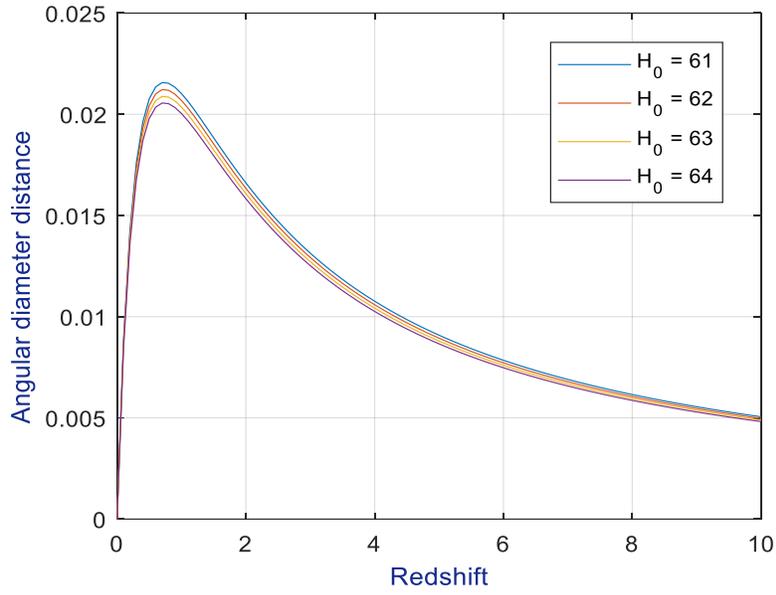

Fig.12 Conduct of angular diameter distance $d_A$ with respect to redshift $z$

with $q = -0.38$, $c' = -1$ and for the values of $H_0$ =61, 62, 63, 64.

The graph of $d_A$ verses $z$ is presented by figure 12 by taking the values $q = -0.38$, $c' = -1$ and different values of $H_0$.

### 4.5 The distance modulus:

The distance modulus $\mu(z)$ is given by

$$\mu(z) = 5\log d_L + 25 \tag{47}$$

$$\mu(z) = 5\log\left\{\frac{9}{qc'^2} H_0^{-1}\left[1-(1+z)^{-2}\right](1+z)\right\} + 25 \tag{48}$$

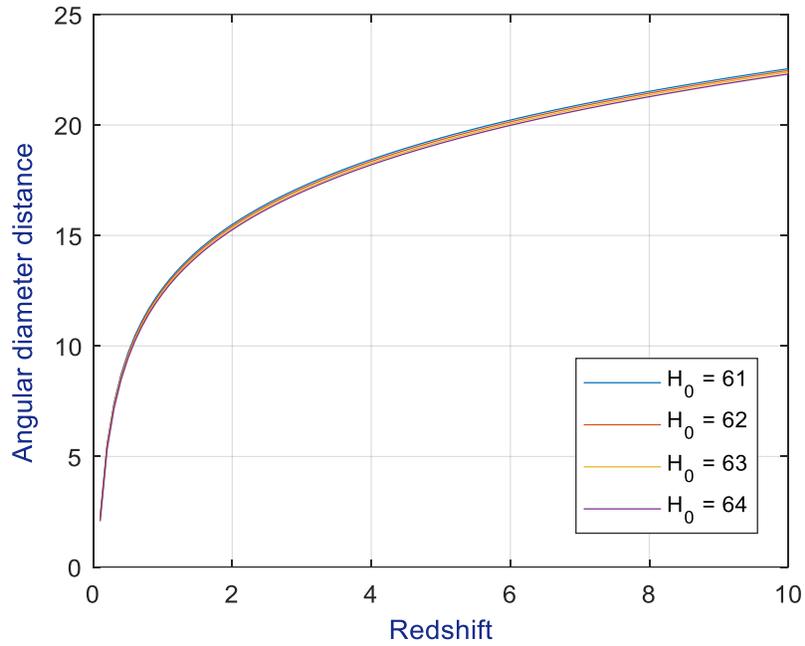

Fig.13 Conduct of distance modulus with respect to $z$

with $q = -0.38$, $c' = -1$ and for the values of $H_0$ =61, 62, 63, 64.

The graph of distance modulus $\mu(z)$ verses redshift $z$ is presented by figure 13 by taking the values $q = -0.38$, $c' = -1$ and different values of $H_0$.

The Hubble parameter $H(z)$ is given by

$$H(z) = H_0 (1+z)^3 \tag{49}$$

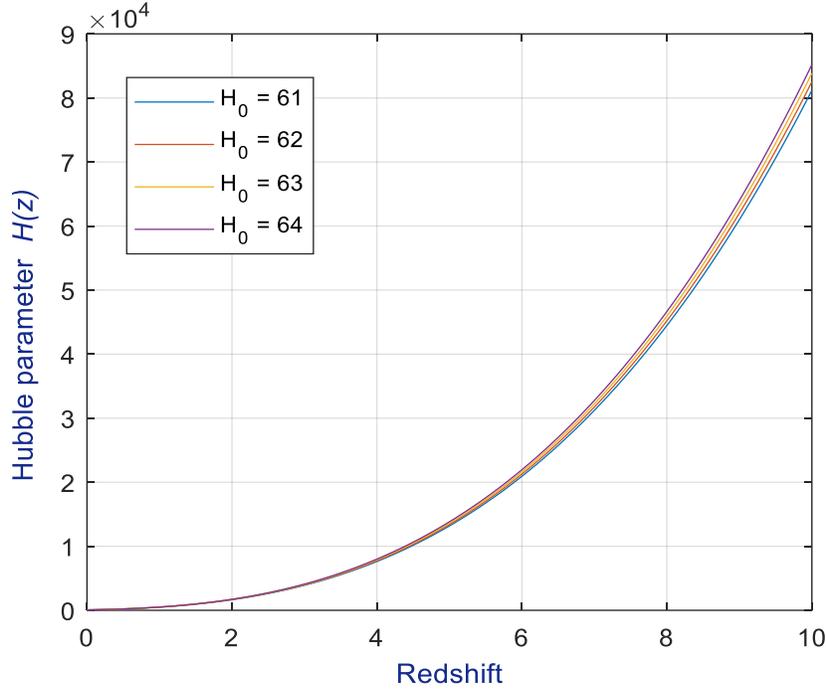

Fig.14 Conduct of Hubble parameter $H(z)$ with respect to redshift $z$

with various values of $H_0$ =61, 62, 63, 64.

The graph of $H(z)$ verses $z$ is presented by figure 14 by taking the different values of $H_0$.

The deceleration parameter $q(z)$ is constant.

**5. Results and discussion-:**

The variation of $\rho_m$ verses $t$ is showed in Figure 1a by choosing the ranges $\eta_1 = 0.1$, $d = 0.5$, $\gamma = 1$ and for the values of $\lambda$ =1, 3, 5, 7, 9. Figure 1b shows the profile of energy density for matter $\rho_m$ against cosmic time by choosing the values $\eta_1 = 0.1$, $d = 0.5$, $\lambda = 3$ and for the values of $\gamma$ =1, 4/3, 5/3, 2. Figure 1c represented behavior of pressure for matter against cosmic time by choosing the values $\eta_1 = 0.1$, $d = 0.5$, $\lambda = 3$ and for the values of $\gamma$ =1, 4/3, 5/3, 2. The graph of energy density for radiation $\rho_r$ verses $t$ is represented in Figure 2 by taking the ranges $\eta_2 = 0.1$, $d = 0.5$, $\gamma = 1$ and different values of $\lambda$. From the Figures 1a, 1b and 2 we observed that, $\rho_m$ and $\rho_r$ are decreasing functions of cosmic time and they approaches towards zero with the evolution of time. From equations (24) and (25), one can show that $\rho_m$ and $\rho_r$ approaches infinity when $T \to 0$. Figures 1a observe that the $\rho_m$ decline as $t$ rises for different interval of $\lambda$ and figure 1b shows that the $\rho_m$ decline as $t$ rises for different interval of $\gamma$. Figure 1b and 1c, respectively shows the energy density and pressure for matter of several universes such as dust ($\gamma = 1$), radiation ($\gamma = 4/3$), hard ($\gamma = 5/3$) and Zel'dovich ($\gamma = 2$) models. From equation (8) and (24) we get pressure for matter and it is constant when $\gamma = 1$. The $\rho_m$, $p_m$ and $\rho_r$ are constants for large values of cosmic time. It has big rip singularity at cosmic time equal to – $d$ and big bang singularity at large values of cosmic time. The intermediate phase is between big bang at $T \to \infty$ and big rip singularity at $T = (t = -d)$.

The graph of density parameter for matter verses cosmic time is given by Figure 3a by choosing the values $\eta_1 = 0.1$, $d = 0.5$, $\gamma = 1$ and for the values of $\lambda =$1, 3, 5, 7, 9. Figure 3b presents the behavior of density parameter for matter against cosmic time by choosing the values $\eta_1 = 0.1$, $d = 0.5$, $\lambda = 3$ and $\gamma =$1, 4/3, 5/3, 2. The variation of density parameter for radiation against cosmic time is represented by Figure 4 by taking $\eta_2 = 0.1$, $d = 0.5$, $\gamma = 1$ and for the values of $\lambda =$1, 3, 5, 7, 9. Figure 3b explores the models such as $\gamma =$1, 4/3, 5/3, 2 of density parameter for radiation. From figure 4 it has been observed that the density parameter for radiation increases with increase in cosmic time increases for the chosen values of $\lambda$ where as in Figure 3b the density parameter for matter for dust ($\gamma = 1$), radiation ($\gamma = 4/3$), hard ($\gamma = 5/3$) and Zel'dovich ($\gamma = 2$) models increases with increase in the cosmic time. From equations (26) and (27) it is seen that the density parameter for matter and density parameter for radiation are infinite when $T \to 0$ and constant for big range of $t$. The density parameter for matter and radiation have big rip singularity at cosmic time equal to – $d$ and big bang singularity at large value of cosmic time.

Figure 5a represented the behavior total density parameter against cosmic time by setting the values $\eta_1 = 0.1$, $\eta_2 = 0.1$, $d = 0.5$, $\gamma = 1$ and for the values of $\lambda =$1, 3, 5, 7, 9. The variation of total density parameter verses cosmic time is shows in Figure 5b by taking the values $\eta_1 = 0.1$, $\eta_2 = 0.1$, $d = 0.5$, $\lambda = 3$ and different values of $\gamma$. Figure 5b the gives total density parameters for different universes such as $\gamma =$1, 4/3, 5/3, 2. The total density parameter observed for the $\gamma =$1, 4/3, 5/3, 2 increases with increase in cosmic time. The total density parameter increases with increase in cosmic time for different interval of $\lambda$ in figure 5a and for different values of $\gamma$ in figure 5b.

The variation of scalar expansion against $t$ is represented in Figure 6 by choosing the value $d = 0.5$. The graph of shear scalar against time by choosing the values $d = 0.5$ and $k = 2.5$ is shown in Figure 7. We observed that Figures 7 and 6 presents that the scalar expansion and shear scalar are decreasing functions of time and they approaches towards zero with the evolution of time. Equations (30) and (31) are displays that the scalar expansion and shear scalar are divergent at the initial stage and vanishes for $T \to \infty$. From figures 7 and 6, the $\theta$ and $\sigma$ decreases as $t$ increases. When $T \to 0$, they approaches towards infinity and have singularity at $T \to \infty$ and $T = (t = -d)$. Equations (30) and (31) shows the universe is expanding and shearing.

The profile of spatial volume verses cosmic time is represented byfigure 8 by taking the values $d = 0.5$ and $b' = 3.1$. The spatial volume is an increasing function of time and it approaches to infinity with the evolution of time. Equation (33) represents that the spatial volume is diverges for $T \to \infty$. It stops at $T \to 0$. From figure 8, the spatial volume is increasing with respect to cosmic time.

## 6. Conclusion:

In present paper, we have explored two-fluid cosmological models in $f(R,T)$ theory of gravity. In this paper, we developed a new idea about f(R, T) gravity with the help of two fluids: one fluid is matter field modeling material content of the Universe and other fluid is radiation field modeling the CMB. Two-fluid universe in theory of $f(R,T)$ gravity is expanding and shearing model.

The energy density and pressure for matter with respective t are shows by Figure 1a, 1b and 1c respectively. The energy density for radiation with respective $t$ is represented by Figure 2 by taking the ranges $\eta_2 = 0.1$, $d = 0.5$, $\gamma = 1$ and different values of $\lambda$. We observed from the Figures 1a, 1b and 2, that $\rho_m$ and $\rho_r$ are decreasing function of cosmic time and approaches towards zero with the evolution of time. From equation (8) and (24) we get pressure for matter and it has constant when $\gamma = 1$.

Figure 1b, 1c, 3b and 5b shows that the energy density, pressure, density parameter and total density parameter for matter presented different universe such as $\gamma$ =1, 4/3, 5/3, 2. We noted from Figure 1b and 1c, that energy density and pressure for matter for dust ($\gamma = 1$), radiation ($\gamma = 4/3$), hard ($\gamma = 5/3$) and Zel'dovich ($\gamma = 2$) models are lessening against $t$. Figure 3b and 5b shows the density parameter and total density parameter for matter of the universes $\gamma$ =1, 4/3, 5/3, 2are increasing against $t$.

We observed that Figures 7 and 6 presented the scalar expansion and shear scalar are decreasing functions of time and approaches towards zero with the evolution of time. Equations (30) and (31) are displays that the scalar expansion and shear scalar are divergent at the initial stage and vanishes for $T \to \infty$. We noted from Figures 7 and 6 that the scalar expansion and shear scalar decreases as cosmic time increases. It has big rip singularity at cosmic time equal to $-d$ and big bang singularity at large value of cosmic time. The transitional stage is between big bang at $T \to \infty$ and big rip singularity at $T = (t = -d)$. Equation (33) represented that the spatial volume is diverge when $T \to \infty$. It is stop at $T \to 0$. Figure 8 shows the spatial volume is increasing with respective cosmic time.

The graph of look-back time $t_L$ verses redshift $z$ is presented by Figure 9 by taking the value $q = -0.38$ and different values of $H_0$. The variation of proper distance $d(z)$, luminosity distance, angular diameter distance and distance modulus $\mu(z)$ verses redshift $z$ is presented by Figure 10, 11, 12 and 13 respectively by setting the values $q = -0.38$, $c' = -1$ and different values of $H_0$. The deceleration parameter $q(z)$ is constant.